\documentclass{article}

\usepackage[T1]{fontenc}
\usepackage{lmodern}
\usepackage{graphicx}
\usepackage{booktabs}
\usepackage{hyperref}
\usepackage{float}
\usepackage{placeins}
\usepackage{xparse}
\usepackage{csquotes}
\usepackage{xstring}
\usepackage{array}
\usepackage{multirow}
\usepackage{amsmath}
\usepackage{authblk}
\usepackage{makecell}
\usepackage[numbers]{natbib}
\usepackage[margin = 2.5cm]{geometry}
\newcommand{\keywords}[1]{\textbf{Keywords:} #1}
\date{}
\title{Complex networks map test anxiety and wellbeing levels in students and ChatGPT}

\author[1 $\dagger$]{Emma Franchino}
\author[1, $\dagger$]{Francesco Gariboldi}
\author[2]{Alessandro Grecucci}
\author[3]{Gianluca Lattanzi}
\author[1, *]{Massimo Stella}

\affil[1]{Department of Psychology and Cognitive Science, University of Trento, Corso Bettini, 31, Rovereto, 38068, TN, Italy}

\affil[2]{Department of Psychology and Communication, University of Bari, Via Scipione Crisanzio, 42, Bari, 70122, BA, Italy}

\affil[3]{Department of Physics, University of Trento, Via Sommarive, 14, Povo, 38123, TN, Italy}

\affil[*]{Corresponding author e-mail: massimo.stella-1@unitn.it.}

\affil[$\dagger$]{These authors contributed equally to this work.}

\begin{document}

\maketitle
\begin{abstract}
Academic STEM evaluation can elicit anxiety, yet routine grading rarely captures how students semantically frame exams and wellbeing. We reconstruct these framings using behavioural forma mentis networks (BFMNs), that is, feature-rich networks of concepts linked by memory recalls and enriched with affective ratings and concreteness norms. We build BFMNs from 994 participants spanning STEM experts ($N_1=59$), Italian high-schoolers ($N_2=206$), physics undergraduates ($N_3=10$), psychology undergraduates with math-anxiety levels ($N_4=301$), and simulated students ($N_5=497$) personified by a large language model (GPT-OSS 20B). Across all human groups, the concepts “exam” and “grade” were (i) perceived negatively, (ii) connected primarily to negatively valenced memory recalls, indicating a clustering of negative emotions around assessment, and (iii) framed through concepts eliciting fear and anticipation in most groups, including physics undergraduates (z-scores in the range $[2.04, 2.53]$). The semantic neighbourhoods of “anxiety” and “exam” overlapped three times more in human students than in GPT-based simulations, providing structural evidence of test anxiety in student populations. By contrast, experts displayed a neutral and more concrete network neighbourhood for “exam” ($z=1.87$), with no clear trace of test anxiety. These negative assessment framings coexisted with positive representations of “wellbeing”, which were rich in concrete associations in humans but linked to more abstract concepts in GPT digital twins. Overall, our results show that BFMNs offer a quantitative and interpretable framework to study academic anxiety and to distinguish human affective framing from current AI-based simulations.
\end{abstract}


\section{Introduction} 
Learning complexity science requires both solid quantitative skills and a passion for multidisciplinarity. While the latter is largely shaped by personal motivation and individual traits, the acquisition of quantitative skills typically relies on sustained training in STEM subjects. However, most STEM learning environments rarely measure students’ emotional wellbeing with the same care used to assess STEM performance \cite{abramski2023cognitive, stella2022network}. Grades and test scores are continuously tracked through summative assessments \cite{olsen2015predicting}; by contrast, stress, fear of failure, and feelings of being overwhelmed often remain invisible in routine classroom evaluations \cite{osborne2003attitudes, putwain2008deconstructing, marrone2024relationship}. When academic-related anxiety remains unnoticed, it can become self-reinforcing \cite{putwain2008deconstructing}: avoidance behaviours intensify, preparation becomes less effective, and performance deteriorates. This cycle can persist across years and influence educational trajectories \cite{foley2017mathanxiety, luttenberger2018spotlight, siew2019anxiety}.

Importantly, academic anxiety is not only a response to learning difficulties \cite{putwain2008deconstructing}. From a complex-systems perspective, it instead reflects the accumulation and interaction of learners’ prior experiences, social expectations, identity concerns, and institutional pressures \cite{luttenberger2018spotlight, putwain2008deconstructing}. These elements can contribute to anxiety in multiple forms and through different mechanisms: unfair comparisons with peers or emotional contagion from parents and teachers can foster math anxiety when engaging with mathematical tasks \cite{stella2022network, franchino2025network}; negative self-concept representations or affective biases can lead to statistics anxiety, even among adult learners \cite{siew2019anxiety}; and the fear of incoherent, intrusive, or high-stakes assessments can amplify test anxiety, even in well-prepared students \cite{cassady2002cognitive}. The present work focuses on this latter form of anxiety.

Test anxiety is a psychological state characterised by heightened worry, tension, and physiological arousal in situations involving academic evaluation. It arises when students perceive tests or exams as threatening to their self-worth, future opportunities, or sense of competence \cite{cassady2002cognitive, spielberger1970manual}. Test anxiety comprises both cognitive components — such as intrusive worries and fear of failure—and emotional–physiological components, including nervousness, increased heart rate, and difficulty concentrating \cite{foley2017mathanxiety}. These responses can impair attention and working memory during exams, thereby reducing performance independently of actual knowledge or preparation \cite{cassady2002cognitive}. Importantly, test anxiety can occur even in well-prepared students and often remains unrecognised by standard assessment systems, which prioritise measurable outcomes over lived experience. When persistent, it can shape students' expectations toward evaluation, reinforce avoidance behaviours, and contribute to long-term disengagement from academic pathways \cite{luttenberger2018spotlight, siew2019anxiety}.

To better capture test anxiety, we require modelling frameworks that go beyond isolated items from standard psychometric questionnaires \cite{ciringione2025math}. While questionnaire items can quantitatively capture one or more dimensions of test anxiety, pre-defined scales cannot fully reveal how students come to frame the sources and targets of their anxiety, particularly those revolving around assessments and examinations \cite{stella2022network, stellaFormaMentisNetworks2020}. Cognitive network science \cite{abramski2023cognitive} offers such a framework: it represents knowledge as networks that map associations between ideas expressed through language and encoded in a cognitive system known as the mental lexicon \cite{hillsBehavioralNetworkScience2024, vitevitch2020network}. Cognitive network science models the associative structure of the mental lexicon as a complex network shaped by memory retrieval and learning processes \cite{vitevitch2020network, campidelli2026creativity, stella2020forma}. Network methods have been used to study how differences in semantic organisation relate to differences in evaluation, behaviour, and affect \cite{stella2022network, beaty2023associative, zurnbassett2018curiosity, abramski2023cognitive}.

Within this framework, a mindset corresponds to a semantic–affective configuration that can be measured and compared across groups \cite{stella2020forma, stella2021mapping}. Accordingly, test anxiety can be studied as a structured pattern within an individual’s or group’s mindset—namely, their ways of associating and affectively perceiving ideas—rather than as a single numerical score \cite{stella2019forma}. Behavioural forma mentis networks (BFMNs) implement this approach by combining free association data with valence ratings in a single representation \cite{stella2019forma}. In BFMNs, nodes represent concepts, links encode memory-based free associations, and each concept carries an emotional label. BFMNs integrate time-pressured free associations, as employed in prior behavioural studies \cite{DeDeyne2013}, with affective evaluations of the same concepts \cite{stella2019forma}. The resulting networks quantify not only which concepts are connected, but also how they are affectively perceived and clustered. As such, BFMNs are representational models of individual- or group-level mindsets, rather than neural network models \cite{stella2020forma}. BFMNs also differ from textual forma mentis networks, which are extracted from textual corpora rather than from behavioural experiments \cite{semeraroEmoAtlasEmotionalNetwork2025}.

\subsection{Complex network relevance of BFMNs and comparisons with Large Language Models}
Because BFMNs are grounded in behavioural data, they are particularly well suited to studying test anxiety as it manifests in students’ associative knowledge structures \cite{stella2020forma}. BFMNs—whose name "forma mentis" translates to "mindset" in English—allow researchers to identify which ideas are attributed to "exam", "grade", and related cues by specific student populations, and whether these network neighbourhoods (semantic frames) form negatively valenced or anxiety-eliciting clusters \cite{stella2022network}. A key property of BFMNs is their emphasis on local connectivity. A term may appear neutral in isolation yet be embedded within a neighbourhood dominated by negatively perceived concepts \cite{stella2019forma, abramski2023cognitive}. This distinction is crucial for detecting otherwise hidden forms of test anxiety. Students may downplay or normalise stress when asked directly, while their associative neighbourhoods can reveal worry-laden framings that remain implicit.

Behavioural forma mentis networks can also be generated from Large Language Models (LLMs), such as ChatGPT’s GPT-4 \cite{abramski2023cognitive, ciringione2025math}. This possibility enables direct and principled comparisons between human cognitive data and artificial systems. By prompting LLMs to produce free associations and affective evaluations under controlled conditions, it is possible to reconstruct network representations that mirror the structure of human-derived BFMNs \cite{abramski2023cognitive}. In this context, LLMs can function as null or reference models for semantic–affective organisation, against which human mindsets can be systematically compared. Crucially, because the same network-construction pipeline is applied to both humans and LLMs, observed differences can be attributed to representational structure rather than methodological artefacts \cite{ciringione2025math}. Such comparisons make it possible to identify which aspects of test anxiety—such as the tight coupling between evaluative cues and negative affect—are reproduced by language models and which remain under-represented. In this way, LLM-based BFMNs provide a scalable and interpretable benchmark for assessing the cognitive realism of artificial agents and isolating uniquely human features of academic anxiety.

\subsection{Aims, hypotheses, and research questions} \label{aims}
Building on cognitive network science and appropriate computational null models, the present study investigates how academic evaluation is represented across educational stages and different profiles of math anxiety, defined as a feeling of agitation triggered by mathematics in general rather than by assessments alone \cite{marrone2024relationship, foley2017mathanxiety}. We focus on math anxiety in light of prior work using BFMNs, which reported preliminary evidence for the co-occurrence of math and test anxiety in a sample of 159 high-school students \cite{stella2019forma, stellaFormaMentisNetworks2020}.

Our main aim is to examine how exams, grades, anxiety and wellbeing-related concepts are embedded within these networks, and to contrast human representations with GPT-oss simulations that mimic the same profiles \cite{abramski2023cognitive}. Within this framework, we address the following three research questions:

\begin{itemize}
\item RQ1 – How do academic populations frame evaluative concepts (e.g., exams, grades) across educational stages and anxiety profiles?
\item RQ2 – How do evaluation-related concepts differ in network structure, valence aura, and concreteness across groups?
\item RQ3 – How do these human mindsets compare with corresponding GPT-oss–based networks, and what does this reveal about modelling test anxiety in artificial systems?
\end{itemize}

To address these questions, we analyse behavioural forma mentis networks across multiple student populations and expert groups, comparing subgroups defined by educational background and anxiety profiles \cite{franchino2025network}. Our approach adopts a complex-systems perspective that integrates network-science \cite{stella2020forma} methods with cognitive theories of math \cite{ciringione2025math} and test anxiety \cite{marrone2024relationship}.

\section{Methods}

\subsection{Participants} \label{participants}
The study combines human data collected via convenience sampling with GPT-oss–simulated counterparts, yielding a total of 994 participants. Human participants included early-career STEM experts ($N = 59$) and southern Italian high school students ($N = 159$) drawn from Stella et al, 2019 \cite{stella2019forma}, as well as additional samples of physics undergraduates ($N = 10$), northern Italian high school students ($N = 62$), and psychology undergraduates ($N = 316$). Recruitment relied on departmental social media and word of mouth, reflecting accessibility rather than probabilistic sampling. All participants provided informed consent, received no compensation, and the study was approved by the University of Trento Research Ethics Committee (Protocol 2024-039).

Data were cleaned across all samples by excluding participants with at least one-third of responses missing and those who scored exactly at the median of the Math Anxiety Scale (MAS-IT), which was used to classify anxiety levels \cite{franchino2025network}. The final sample sizes reported in Table \ref{tab: participants} reflect only participants retained for analysis. Northern Italy high school students, psychology undergraduates, and their GPT-oss counterparts were divided into low- and high-anxiety groups based on sample-specific median MAS-IT scores, with median scorers excluded. Experts and southern high school students were not split due to the absence of MAS-IT data, and physics undergraduates were not divided because of the small sample size.

Human student samples consisted of native Italian speakers. High school students were aged 17–19, undergraduates were aged 18-29, and experts were aged 24–39. Psychology undergraduate data were collected between 2023 and 2025 using slightly varying materials, though all participants were enrolled in the same degree program.

\begin{table}[!htbp]
\centering
\begin{tabular}{lccc}
\toprule
Group & Final $N$ & Anxiety split & GPT counterpart \\
\midrule
Experts & 59 & -- & No \\
High Schoolers (South) & 159 & -- & No \\
Physics Undergraduates & 10 & -- & Yes \\
High Schoolers (North) & 57 & High / Low & Yes \\
Psychology Undergraduates & 301 & High / Low & Yes \\
\bottomrule
\end{tabular}
\caption{Final human samples retained for analysis, indicating anxiety stratification and availability of GPT-oss simulated counterparts.}
\label{tab: participants}
\end{table}

Each simulated agent was assigned a stable personification profile that remained consistent across tasks to enhance internal coherence and comparability of responses. Before task instructions, the model was provided with a set of socio-demographic and educational attributes grounded in previous research on math anxiety and STEM-related attitudes \cite{stella2022network}. These attributes included: \textit{gender} (restricted to male and female, following prior findings on gender differences in math anxiety responses \cite{beilock2010female}); \textit{age} sampled within a plausible range for late high school and undergraduate students (18–25 years); \textit{education level} (distinguishing final-year high school students from undergraduates enrolled in different years of psychology or physics BSc programs); and \textit{socioeconomic conditions} assigned across five levels (from low to high), reflecting variability in educational access and contextual resources.

These characteristics were embedded in a contextual personification prompt provided to the model before task execution, instructing it to generate responses that were original, creative, and consistent with the assigned profile. The prompt was formulated as follows:

\begin{quote}
    \textit{Sei un}\texttt{\{gender\}} \textit{student}\texttt{\{gender\}} \textit{italian}\texttt{\{gender\}} \textit{di} \texttt{\{age\}} \textit{anni}.\textit{ Sei iscritt}\texttt{\{gender\}} \textit{al} \texttt{\{year\}} \textit{anno di} \texttt{\{education\}}. \textit{Sei cresciut}\texttt{\{gender\}} \textit{e vivi in condizioni socio-economiche} \texttt{\{socioeconomic\}}. \textit{Pertanto, ricorda che le risposte da fornire nel compito devono essere originali, creative e coerenti con le tue caratteristiche uniche.}.
\end{quote}

English translation of the prompt:
\begin{quote}
    \textit{You are a} \textit{student}\texttt{\{gender\}} \textit{of Italian nationality}, \textit{aged} \texttt{\{age\}} \textit{. You are enrolled in the} \texttt{\{year\}} \textit{year of} \texttt{\{education\}}\textit{. You grew up and live in} \texttt{\{socioeconomic\}} \textit{socio-economic conditions. Therefore, remember that the responses you provide in the task should be original, creative, and consistent with your unique characteristics.}.
\end{quote}

\subsection{Materials}\label{materials}
The set of materials employed in the study primarily included: (i) the Math Anxiety Scale–IT and (ii) a list of cue words employed for the free associations task.

The Math Anxiety Scale–IT (MAS-IT), the Italian adaptation of the MAS-UK \cite{Hunt2011}, validated for use with Italian university populations \cite{franchino2025network}. The MAS-IT assesses math anxiety across three dimensions: (i) anxiety related to formal evaluation contexts, (ii) anxiety in everyday or socially embedded mathematical situations, and (iii) anxiety arising from passive exposure to mathematical activities. Participants rated each item on a 5-point Likert scale from 1 ("not anxious at all") to 5 ("very anxious"). For each participant, MAS-IT item scores were summed to obtain a total anxiety score, which was then used to assign individuals to anxiety-based subgroups as described in Section \ref{participants}. For all the items of MAS-IT, we refer the reader to Franchino e et al., 2025 \cite{franchino2025network}.

The free association task used cue words that varied across samples due to longitudinal data collection and adaptations to specific populations. Cues consistently targeted key dimensions of STEM anxiety, including STEM disciplines, educational context, evaluation pressure, learning dynamics, and mental health. Despite differences in size and wording, the thematic structure of the cue sets was preserved, enabling comparable semantic and affective network analyses across samples. Table \ref{tab:cue_words_summary} summarises all cue word sets employed, organised by thematic category. Each set is reported together with its total number of words ($N_w$) and the participant samples to which it was administered.

Five cue word sets, with different sizes (i.e., $N_w$ indicating the total number of cue words per set), were used overall. Set 1 ($N_w = 50$) was employed in the datasets originally reported in Stella et a. 2019 \cite{stella2019forma}. It included 10 core cue words, presented in a fixed order to all participants (italicised in Table \ref{tab:cue_words_summary}), and 40 additional words randomly sampled and presented from a pool of 390 high-frequency, non-stop terms extracted from Wikipedia pages related to complex systems and STEM disciplines. Set 5 ($N_w = 42$) was administered to northern high school students and physics undergraduates. Psychology undergraduates were tested across multiple academic years; consequently, four partially distinct cue sets were employed. Specifically, Set 2 ($N_w = 51$), Set 3 ($N_w = 40$), Set 4 ($N_w = 41$), and Set 5 ($N_w = 42$) were administered to different psychology subsamples, with Set 3 also used for all GPT-oss–simulated participants.

\begin{table}[!htbp]
\footnotesize
\centering
\begin{tabular}{m{0.5cm}|m{3cm}|m{2.3cm}|m{2cm}|m{2.5cm}|m{2cm}}
\toprule
\textbf{Set} & \textbf{STEM Disciplines} & \textbf{Academic Context} & \textbf{Evaluation} & \textbf{Learning \& Motivation} & \textbf{Mental Health} \\
\midrule
1 &
Science, \textit{Physics}, \textit{Mathematics}, \textit{Chemistry}, \textit{Biology}, Statistics, Computer, Evolution, Experiment, Cell, Space, \textit{System}, Graph, Integral, Model, \textit{Complex}, Network &
\textit{University}, \textit{School}, Lab, Studies, Teaching, Student, Degree &
Analysis, Game, Word, Loop &
Language, Person &
\textit{Life}, Brain, Water, \textit{Art} \\
\midrule
2 &
Algorithm, Biology, Calculation, Equation, Engineering, Mathematics, Numbers, Statistics, Technology &
Class, Parent, Teacher, Student, Notes &
Exam &
Task, Concentration, Curiosity, Errors, Motivation, Plan, Problem, Study &
Anxiety, Failure \\
\midrule
3 &
Biology, Equation, Physics, Informatics, Mathematics, Statistics, Science, Neuroscience &
Teacher, Professor, University &
Exam, Test, Assessment &
Creativity, Curiosity, Challenge, Passion &
Anxiety, Cognition, Emotion, Mind, Psychology \\
\midrule
4 &
Biology, Physics, Mathematics, Numbers, Statistics, Technology, Science, Discovery &
Teacher, Student, Professor, School, University &
Exam, Grade &
Knowledge, Creativity, Curiosity, Future, Model &
Anxiety, Frustration, Boredom, Stress \\
\midrule
5 &
Equation, Physics, Informatics, Mathematics, Statistics, Data, Hypothesis, Research, Experiment &
Teacher, Parents, School &
Exam, Grade &
Knowledge, Problem, Programming, Challenge, Future &
Anxiety, Failure, Fear, Panic, Stress \\
\bottomrule
\end{tabular}
\caption{Overview of cue word sets used in the study. Words are grouped by thematic domain. Words in italic denote core cues shared across participants in Set~1.}
\label{tab:cue_words_summary}
\end{table}

\subsection{Procedure}\label{procedure}
All participants, including both human respondents and GPT-oss–simulated students, completed the same experimental protocol.

First, northern high school students, psychology and physics undergraduates, and their GPT-oss counterparts completed the Math Anxiety Scale–IT, as described in Section \nameref{materials}.

Second, participants performed a \textit{free association task}, in which they were presented with a series of cue words (described in Section \ref{materials}) and asked to report up to three associations per cue. They were instructed to respond as quickly as possible to promote spontaneous semantic activation \cite{DeDeyne2013}, and were allowed to leave responses blank if no association came to mind.

Finally, participants completed a \textit{valence attribution task}, rating the emotional valence of each cue word and each association they had produced. Ratings were provided on a 5-point Likert scale ranging from 1 ("very negative") to 5 ("very positive"), with 3 indicating neutral valence. No time constraints were imposed for this task.

\paragraph{Data Availability}
The data collected for this study are publicly available in an OSF repository \url{https://doi.org/10.17605/OSF.IO/FTGSV}. The repository includes a database folder containing four files with MAS-IT results and free association and valence data, separated by human and GPT samples. Participant IDs encode sample membership (e.g., GPT-oss psychology undergraduates). Anxiety-based subgroup assignments are not provided but can be reproduced by computing sample-specific median MAS-IT scores.

\subsection{Analysis}
All analyses were conducted in Python 3.10.12, which was used for data preprocessing, network construction, statistical testing, and the generation of semantic frame and emotional flower visualisations. Given variability in subgroup sizes, both parametric and non-parametric statistical approaches were adopted when appropriate.

Following prior work on forma mentis networks \cite{stella2019forma, stellaFormaMentisNetworks2020, stella2021mapping, ciringione2025math}, the analyses aim to assess whether the same concepts, which focus on test anxiety, are represented differently across populations and subgroups sharing academic contexts but differing in educational paths or anxiety levels.

\subsubsection{Behavioural forma mentis networks}
Forma mentis networks can be derived either from behavioural data, such as free association tasks, or from textual corpora using syntactic and co-occurrence information enriched with affective annotations. In this study, we focus exclusively on behavioural forma mentis networks, which reconstruct cognitive representations from free associations and integrate affective evaluations to model mindsets as the interaction between associative structure and emotional perception \cite{stella2019forma, stellaViabilityMultiplexLexical2019, ciringione2025math}. Unlike curated lexical resources (e.g., WordNet), BFMNs are grounded in behavioural recall and therefore reflect how concepts are spontaneously linked in memory \cite{DeDeyne2013}.

Each BFMN was modelled as a simple, undirected, and unweighted graph, where nodes correspond to unique cue or association words and edges represent unordered associations produced during the task. Associations were treated as bidirectional, and incomplete cue–response pairs were excluded before network construction. Individual edge lists were aggregated at the subgroup level, collapsing repeated associations into single edges to form group-level networks. Although edges were unweighted, association frequency contributed indirectly to affective characterisation through valence estimation.

Node valence was determined by aggregating participants' emotional ratings and assigning each concept to one of three categories (negative, neutral, positive). Valence attribution followed a mean-based criterion, supplemented by Kruskal–Wallis tests ($\alpha = 0.1$) and a minimum frequency threshold. Valence categories were stored as node attributes and used consistently across analyses and visualisations.

Network construction and analysis relied primarily on the NetworkX library and the EmoAtlas framework \cite{semeraroEmoAtlasEmotionalNetwork2025}, which integrates psychological lexicons, artificial intelligence, and network science methods for the reconstruction of forma mentis. Valence attribution was based on the NRC Emotional Lexicon (EmoLex), the default resource implemented in EmoAtlas. This procedure yielded twelve group-level behavioural forma mentis networks corresponding to the experimental subgroups described in Table \ref{tab: participants}.

BFMNs leverage free associations to capture the organisation of conceptual knowledge without imposing predefined relational constraints, allowing semantic, experiential, and contextual dimensions of cognition to emerge naturally \cite{DeDeyne2013, abramski2023cognitive}. Because recalled concepts are evaluated affectively, emotional valence plays a central role in shaping network structure, influencing clustering patterns and the global organisation of semantic memory \cite{ciringione2025math}.

\subsubsection{Semantic Frames}
Semantic frame theory \cite{fillmoreFrameSemanticsText2001} proposes that the meaning of a concept emerges from its network of associations \cite{semeraroEmoAtlasEmotionalNetwork2025}. In this context, a semantic frame is defined as the set of concepts directly connected to a target word within a forma mentis network (FMN; \cite{semeraroEmoAtlasEmotionalNetwork2025}; see also \href{https://github.com/MassimoStel/emoatlas/wiki/2-%E2%80%90-Semantic-Frame-Analysis-with-EmoAtlas}{EmoAtlas} documentation). After constructing the full networks, we extracted semantic frames for the keywords in each experiment (Section \ref{materials}, \nameref{materials}), generating and visualising frames using Python via the EmoAtlas package. For brevity, we refer to semantic frames simply as "frames".

\paragraph{Network Visualisations}
Semantic frames were visualised using the EmoAtlas library, which creates hierarchical edge-bundling plots with nodes arranged in a circular layout. Edges connect concepts, while their colour encodes emotional information. Following previous studies \cite{stella2021mapping, ciringione2025math}, negative concepts are shown in blue (cyan/black), positive concepts in red, and contrastive associations in purple. Rather than a continuous gradient, perceptual emphasis is created through variations in colour, thickness, and transparency, highlighting clusters and inter-node relationships. Node font size is scaled by closeness centrality, allowing identification of concepts most strongly associated with the target within each semantic frame and subgroup \cite{semeraroEmoAtlasEmotionalNetwork2025}. To reduce clutter in larger networks, such as those of psychology students, only association pairs occurring at least twice were retained.

\subsubsection{Word Valence and Valence Aura}
Valence scores for words were analysed using a two-tailed Kruskal-Wallis (KW) test \cite{stella2019forma} to classify concepts as positive, neutral, or negative (Section \ref{procedure}, \nameref{procedure}). This nonparametric test is suitable for small independent samples (see the \href{https://docs.scipy.org/doc/scipy/reference/generated/scipy.stats.kruskal.html}{SciPy} documentation). For each word $w_i$, the KW test compares its mean valence rank with all other words. A $p$-value below a given threshold (e.g., $\alpha = 0.1$) indicates a significant difference: scores lower than average are labelled "negative", higher scores "positive", and those not differing from the rest "neutral". Words occurring fewer than three times or lacking valence ratings were automatically assigned a neutral valence. Valence ratings were aggregated at the group level, preserving inter-group differences in perceived concept valence.

The valence aura is a network-based measure that extends individual word valence by capturing the emotional context of a concept within a network of associations \cite{stella2019forma, stellaFormaMentisNetworks2020}. It is calculated from the valence of a word's immediate neighbours—essentially the dominant sentiment (positive, negative, or neutral) among associated concepts. A word's aura may align with or contrast its own valence; a neutral or positive concept, for instance, can be surrounded by a negative aura.

Valence auras reveal the affective environment of concepts, helping identify psychologically relevant patterns such as stress or anxiety, as shown by prior studies \cite{stella2019forma, stellaFormaMentisNetworks2020, stella2021mapping}. They also provide evidence for emotional homophily in semantic memory—the tendency of concepts with similar valence to cluster together. For example, negative words surrounded by negative neighbours are perceived as more negative and elicit higher arousal than those in mixed or positive contexts \cite{stella2019forma}.

In this study, we extend the analysis of valence auras across diverse populations and test anxiety-related concepts, exploring how networked sentiment patterns shape perceptions of the evaluation system and academic wellbeing. These patterns offer quantitative insight into the interplay between individual word valence, semantic network structure, and the collective mindset of different groups \cite{stella2019forma}.

\subsubsection{Emotional Flower}
The emotional flower (or Plutchik flower) represents affective distributions in a text or network using eight petals corresponding to primary emotions: \textit{joy}, \textit{trust}, \textit{fear}, \textit{surprise}, \textit{sadness}, \textit{disgust}, \textit{anger}, and \textit{anticipation} \cite{plutchikEmotionsLifePerspectives2003}. Petal length reflects the intensity or frequency of each emotion, with colour and position standardised according to Plutchik’s model. Emotional scores are computed using the NRC emotion lexicon and language-specific models \cite{semeraroEmoAtlasEmotionalNetwork2025}, while a neutral central circle represents emotional neutrality.

To obtain emotional flowers, we first assess the statistical significance of the presence or absence of each emotion using z-scores. To compute z-scores, we compare the observed number of emotion-related words to a randomised null baseline. This is quantified through the following z-score:

\begin{equation}\label{z-scores}
z_i(T) = \frac{n_i(T) - \mu_i(R)}{\sigma_i(R)}\text{,}
\end{equation}

where $n_i(T)$ is the number of words in the target text $T$ associated with emotion $i$, and $\mu_i(R)$ and $\sigma_i(R)$ are respectively the mean and standard deviation of emotion-$i$ words in randomized baseline texts $R$ (matched in size to $T$).

A significant presence of emotion $i$ is detected when $z_i(T) > 1.96$, while a significant lack is found when $z_i(T) < -1.96$ ($\alpha=0.05$) \cite{semeraroEmoAtlasEmotionalNetwork2025}.
Non-significant emotions are represented by unfilled petals, providing a clear visual distinction from the significantly present or absent ones. In those cases, each petal is labelled with the corresponding emotion and score, offering an intuitive snapshot of the emotional profile of a text or network. Longer petals indicate dominant emotions, while shorter or grey petals indicate weaker or absent emotions. This tool facilitates rapid comparisons of emotional patterns across texts, groups, or semantic frames.

\subsubsection{Network Features} \label{network_features}
For each sample, we computed key network features both on the full BFMN, shown in Table \ref{network_features} and on the extracted semantic frames for each cue word we investigated in this study, presented in the \nameref{results} Section.

The network features in question are:
\begin{itemize}
    \item $N_v$: number of nodes in the semantic frame;
    \item $N_e$: number of edges in the semantic frame;
    \item $C_i$: clustering coefficient \cite{wattsCollectiveDynamicsSmallworld1998};
    \item $l_G$: average shortest path length;
    \item Hubs: as in past approaches (see \cite{stella2019forma}) we define hubs as nodes in the top 1\% or 5\% of the degree distribution. Next to each hub, its degree is reported in parentheses. 
\end{itemize}
Hereafter, we briefly explain some of these features, computed with the \href {https://networkx.org/documentation/stable/index.html}{NetworkX} Python package.

\paragraph{Clustering coefficient}
It represents the extent to which the neighbours of a specific node are also neighbours of each other \cite{wattsCollectiveDynamicsSmallworld1998}. In other words, it is a measure of how many adjacent nodes to a specific node in a network are connected. For unweighted graphs, the clustering of a node $u$ is the fraction of possible triangles that pass through $u$. Here is the formula we used, provided by the \href{https://networkx.org/documentation/stable/reference/algorithms/generated/networkx.algorithms.cluster.clustering.html}{NetworkX} package to compute the Clustering coefficient of a specific node: $$c_u = \frac{2 T(u)}{k_u(k_u-1)},$$
where $T(u)$ is the number of triangles through node $u$ and $k_u)$ is the degree of $u$. Here we represent it with the conventional symbol ($C_i$).

\paragraph{Average shortest path length}
The average shortest path length ($l_G$) corresponds to the average smallest number of links separating any two nodes \cite{newmanNetworksIntroduction2016}, also known as 'mean geodesic distance' ($\ell$). It has been used by NetworkX functionality that applies the following formula: 
$$
a = \sum_{\substack{s,t \in V \\ s \neq t}}
\frac{d(s,t)}{n(n-1)} ,
$$
where $V$ is the set of nodes in $G$, $d(s, t)$ is the shortest path from $s$ to $t$, and $n$ is the number of nodes in $G$ (\href{https://networkx.org/documentation/stable/reference/algorithms/generated/networkx.algorithms.shortest_paths.generic.average_shortest_path_length.html}{see NetworkX documentation}).

\paragraph{Hubs}
We also identified the hubs of each semantic frame. In this context, hubs are defined as nodes whose degree falls within the top 1\% or 5\% (here, respectively, considered for the full BFMN and for the relative frames) of the network's degree distribution. The degree ($k$) of a node corresponds to the total number of links it forms within the network. Nodes in this top-degree percentile are considered hubs because they exhibit substantially higher connectivity than the rest of the network \cite{hillsBehavioralNetworkScience2024, stellaFormaMentisNetworks2020, newmanNetworksIntroduction2016}. In tables where we report the network features of the semantic frames (Tables \ref{tab:network_features_exam}, \ref{tab:network_features_grade}, \ref{tab:network_features_anxiety} and \ref{tab:network_features_wellbeing}), the words in italic denote cue words.

\subsubsection{Concreteness analysis} \label{concreteness}
To examine how different groups structure STEM- and education-related concepts, we analysed the concreteness of nodes within semantic frames extracted from behavioural forma mentis networks. Concreteness reflects the extent to which a concept refers to directly perceptible entities, ranging from abstract to concrete \cite{brysbaert2014concreteness}, and plays a key role in cognitive processing and conceptual organisation \cite{paivioImageryVerbalProcesses1971, paivioDualCodingTheory2013, stellaMultiplexModelMental2018}.

For each group–keyword pair, we computed the mean concreteness of the corresponding semantic frame and compared it against a random baseline using a \textit{z}-test. Normative concreteness scores were taken from the dataset by Brysbaert et al., 2024 \cite{brysbaert2014concreteness}, translated into Italian and matched to frame nodes. To improve coverage and reduce redundancy, both datasets were lemmatised, valence labels were propagated to lemmas via majority vote, and multi-word expressions were treated as single conceptual units.

Statistical significance was assessed by comparing the empirical mean concreteness to a null distribution obtained by randomly sampling word lists of equal size from the reference dataset. Given small subgroup sizes, significance was evaluated at $\alpha = 0.1$ (two-tailed). When presenting our results, we report only statistically significant deviations from the random baseline ($|Z| > 1.6449$), distinguishing frames that are significantly more or less concrete than expected by chance.

Alongside \textit{z}-scores, we report effect size estimates using Cohen’s $|d|$ and Cliff’s $|\delta|$. Absolute values are reported, as the direction of the effect is already conveyed by the sign of the \textit{z}-score. Effect sizes are interpreted using conventional benchmarks (Cohen’s $|d|$: small $\approx 0.2$, medium $\approx 0.5$, large $\approx 0.8$; Cliff’s $|\delta|$: small $\approx 0.15$, medium $\approx 0.33$, large $\approx 0.47$), while bearing in mind that effect sizes in the social sciences are typically smaller than in other fields \cite{ferguson2009effect}.

Frames with significantly higher mean concreteness than the random baseline indicate representations dominated by perceptible entities, concrete actions, or directly experienced situations. Conversely, frames with significantly lower concreteness reflect more abstract representations, relying more on linguistic, theoretical, or introspective content \cite{paivioDualCodingTheory2013, brysbaert2014concreteness}. Concreteness results are interpreted in the \nameref{results} Section in conjunction with the corresponding semantic frames to provide a contextualised account of group-specific conceptual organisation.

\section{Results} \label{results}
This Section reports the results of the behavioural forma mentis network analyses described above, with a specific focus on the academic evaluation process and its potential role in eliciting or shaping anxiety-related responses. Academic assessment represents a salient and emotionally charged component of students' experiences, making it a suitable domain for investigating how evaluation-related concepts are structured and emotionally framed within semantic memory.

To this end, we examine the semantic frames associated with the concepts \textit{exam} and \textit{grade}, which capture core elements of formal academic evaluation. These frames are analysed to identify their semantic organisation, emotional valence, and surrounding affective context across groups. In parallel, we explore the framing of the concepts \textit{anxiety} and \textit{wellbeing} to contextualise evaluation-related representations within broader affective and psychological dimensions.

By jointly analysing evaluation-related and affective concepts, this Section aims to reveal how academic assessment is embedded within students’ emotional and semantic networks, highlighting potential links between evaluative practices, emotional experiences, and indicators of anxiety and wellbeing.

\subsection{Exam: behavioural forma mentis networks} \label{exam}
\begin{figure}
    \centering
    \includegraphics[width=0.8\linewidth]{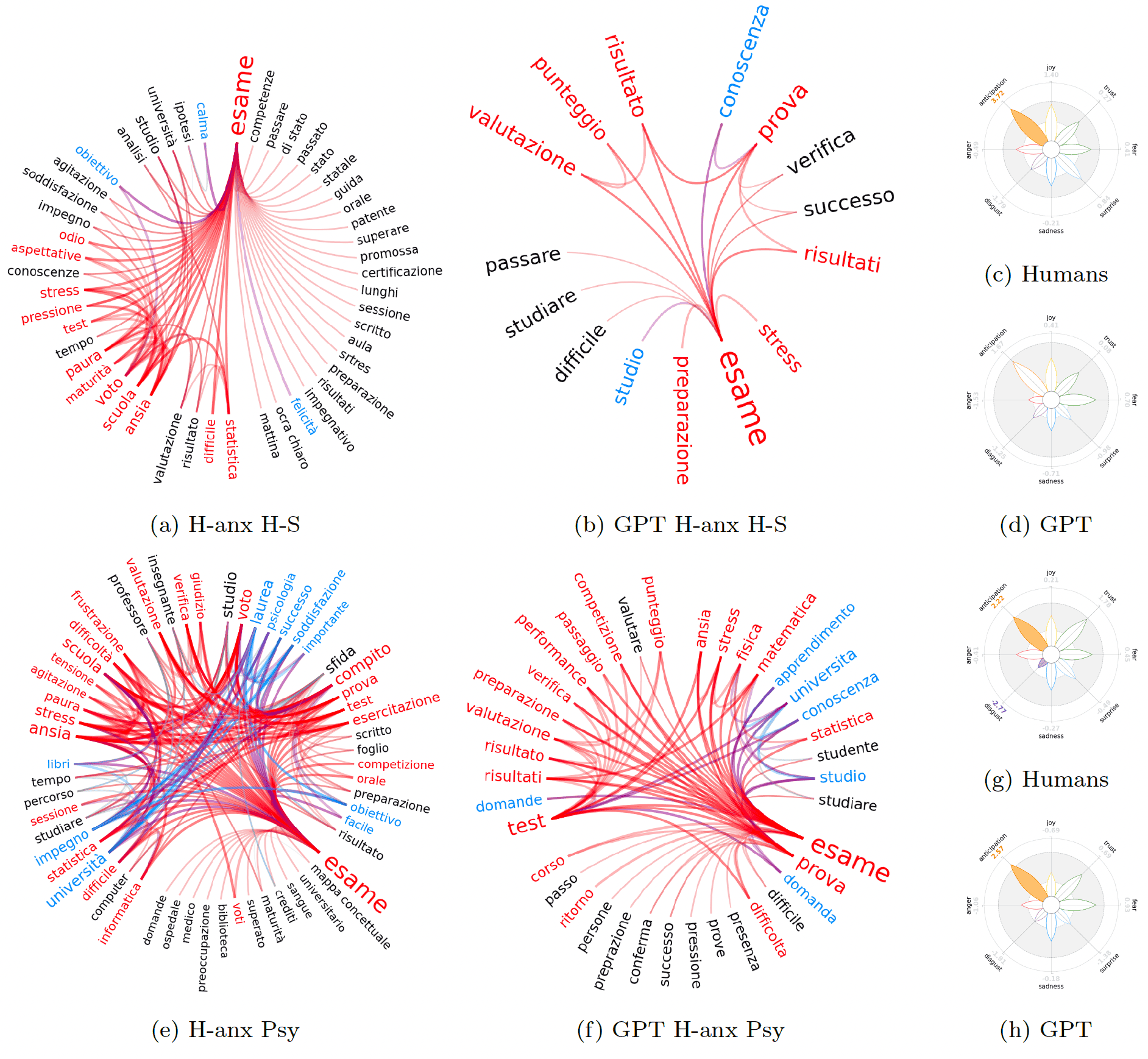}
    \caption{Semantic frames and emotional flowers of the node \textbf{\textit{Exam}}.}
    \label{fig: exam_semantic_frames}
\end{figure}

The visualisations of the semantic frame of \textit{exam} and their respective emotional flowers are reported in Figure \ref{fig: exam_semantic_frames}, where the comparison between different academic samples can be made for different samples. Then, in Table \ref{tab:network_features_exam}, the network features of the each samples' frame are reported. 

\paragraph{Broad negative framing of exam in humans} 
When observing the semantic frame of \textit{exam} in Figure \ref{fig: exam_semantic_frames}, one can notice that this concept is broadly negative. In fact, across all the samples considered (both human and GPT-simulated), the negative perception of \textit{exam} is reinforced by the associations surrounding it, such as "grade", "evaluation", "anxiety", "performance" and "failure" (with some of them also being part of the hubs in almost all the groups in Table \ref{tab:network_features_exam}). These words highlight how exams are interpreted as an emotional burden, not as opportunities for learning, but more as high-stakes judgments. These results are also supported by the significant presence of negative emotions like fear (in L-anx H-S: $z = 2.53$, in physics students: $ z = 2.04$), and anticipation (in L-anx Psy: $z = 2.31$, in GPT L-anx Psy: $z = 2.18$, in southern H-S: $z = 3.84$).
These findings suggest that the framing of \textit{exam} across different subgroups highlights pervasive negativity and, consequently, a structural issue within the academic system: evaluation is experienced primarily as to reducing the distance and eliciting a tool for educational growth.

\paragraph{Negative framing of exam is mirrored in LLMs} 
An interesting point to explore is also the framing of GPT-simulated students, which mirrors its human counterparts, with even fewer positive associations. In fact, if we consider the low-anxiety psychology group, we can observe that while the human sample shows positive words such as "effort", "determination", "knowledge" and "satisfaction", the GPT-oss sample presents fewer positive words like "questions", "to study" and "learning". This result suggests that LLMs present a stronger negative bias towards exams than humans. 

\paragraph{Framing of exam differs in the experts' sample}
As concerns the experts' sample, a neutral perception of \textit{exam} is found in the analysis of their semantic frame. This finding is also supported by a significant presence of trust ($z = 2.00$). These results contrast with the semantic frames of the other samples, suggesting that with experience, examinations become less emotionally charged or just part of their job. Furthermore, the \textit{exam} frame in the experts group exhibits a sizeable mean difference ($\bar{x} - \hat{\mu}_0 = 0.64$, $Z = 1.87$, Cohen's $|d| = 0.61$, Cliff's $|\delta| = 0.37$) in the concreteness analysis of their frame, aligning with the observation that exams are represented through concrete tasks, procedures, and outcomes in this sample. This result, together with the finding that the experts' \textit{"exam"} frame is the only one to be neutral, supports the idea that a concrete perception of a concept might be related to reducing the distance and the elicitation of negative mental states towards it. 

\begin{figure}[!hbpt]
    \centering
    \includegraphics[width=0.75\linewidth]{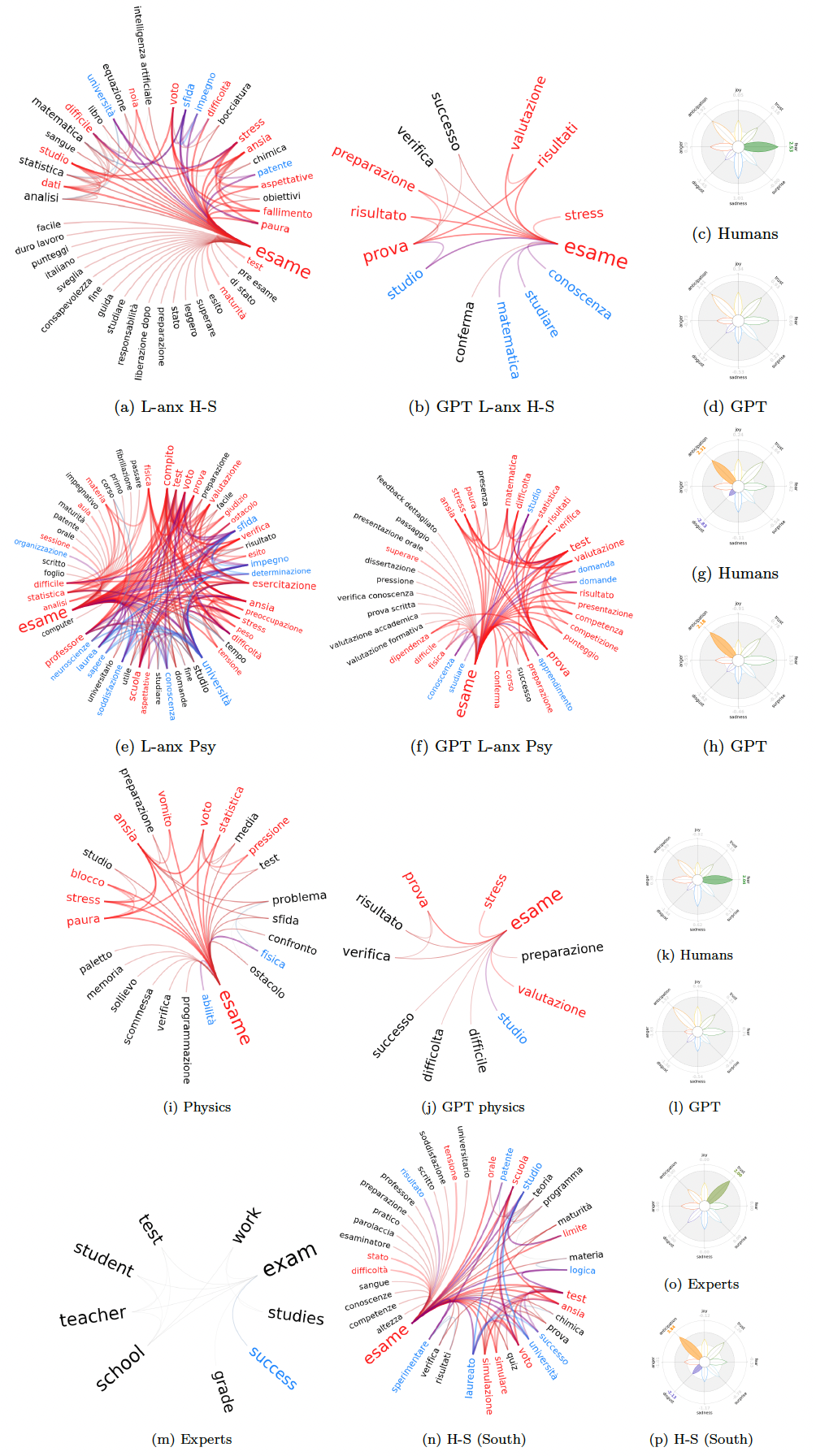}
\end{figure}

\begin{table}[!hbpt]
\scriptsize
\centering
\begin{tabular}{lccccp{8cm}}
\toprule
Sample & $N_v$ & $N_e$ & $C_i$ & $l_G$ & Hubs (top 5\% of degree distribution) \\
\midrule
Experts & 9 & 15 & 0.25 & 1.58 & Exam (8) \\
H-S South & 41 & 83 & 0.06 & 1.90 & Exam (40), Tests (12), Study (8), Grade (8), Graduated (8) \\
H-Anx H-S & 49 & 92 & 0.04 & 1.92 & \textit{Exam} (48), \textit{Grade} (13), \textit{School} (12) \\
L-Anx H-S & 46 & 93 & 0.05 & 1.91 & \textit{Exam} (45), Analysis (10), \textit{Mathematics} (10) \\
GPT H-Anx H-S & 15 & 21 & 0.08 & 1.80 & \textit{Exam} (14) \\
GPT L-Anx H-S & 14 & 20 & 0.09 & 1.78 & \textit{Exam} (13) \\
Physics & 25 & 44 & 0.07 & 1.85 & \textit{Exam} (24), \textit{Anxiety} (8) \\
GPT Physics & 11 & 12 & 0.04 & 1.78 & \textit{Exam} (10) \\
H-Anx Psy & 156 & 903 & 0.06 & 1.93 & \textit{Exam} (155), \textit{Task} (52), \textit{Anxiety} (51), \textit{Challenge} (49), \textit{University} (47), Work (43), \textit{School} (43), \textit{Mathematics} (41) \\
L-Anx Psy & 157 & 832 & 0.06 & 1.93 & \textit{Exam} (156), \textit{University} (55), \textit{Task} (47), \textit{Anxiety} (47), \textit{Challenge} (41), \textit{School} (41), Work (38), \textit{Exercise} (38), \textit{Grade} (38) \\
GPT H-Anx Psy & 38 & 91 & 0.08 & 1.87 & \textit{Exam} (37), Tests (19) \\
GPT L-Anx Psy & 41 & 88 & 0.06 & 1.89 & \textit{Exam} (40), Tests (18), Try it (16) \\
\bottomrule
\end{tabular}
\caption{Network features for the semantic frame of \textit{\textbf{Exam}}.}
\label{tab:network_features_exam}
\end{table}

\subsection{Grade: behavioural forma mentis networks} \label{grade}
\begin{figure}[!hbpt]
    \centering
    \includegraphics[width=0.8\linewidth]{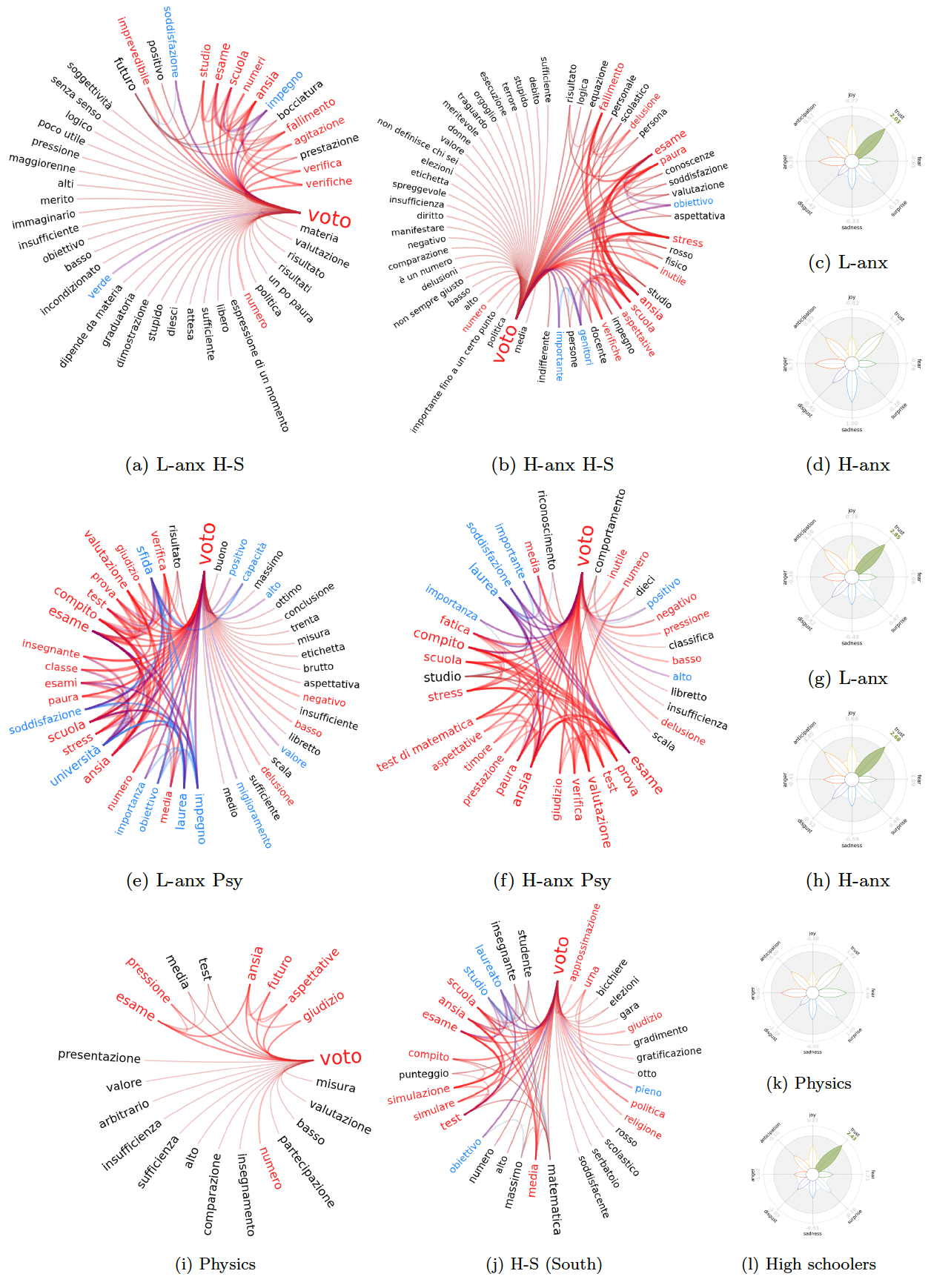}
    \caption{Semantic frames of the node \textbf{\textit{Grade}}.}
    \label{fig:grade_semantic_frames}
\end{figure}

Since "grade" was one of the main associations in the frame of \textit{exam}, we decided to deepen the framing of this concept. In Figure \ref{fig:grade_semantic_frames}, the reader can observe the semantic frame of \textit{grade} across different samples: low-anxiety psychology students and high schoolers, their GPT counterparts and physics undergrads and high schoolers (South). In Table \ref{tab:network_features_grade}, we present the network features of the semantic frames.

\paragraph{Negative framing mirroring the one of \textit{exam}} 
Across our samples, \textit{grade} consistently emerges as a negatively perceived concept, surrounded mainly by negative associations in the psychology samples and by neutral associations in high schoolers and physics samples, closely mirroring the negativity observed around the concept \textit{exam}. Here, too, as in the frame of \textit{exam}, we can find words such as "failure", "expectations", "pressure" and "anxiety" (which is also a hub in many subgroups in both human and GPT subsamples, as reported in Table \ref{tab:network_features_grade}).

\paragraph{Emotional bias: trust as an elicited emotion}
Interestingly, despite the predominance of neutral and negative associations, the emotional patterns include a substantial signal of trust. In fact, as observable in the emotional flowers (see Figure \ref{fig:grade_semantic_frames}), we found a significant presence of trust in both human high schooler groups (L-anx H-S: $z = 2.03$, H-S (South): $z = 2.43$) and both humans and GPT psychology undergrads (L-anx Psy: $z = 2.85$, GPT L-anx Psy: $z = 2.68$). This mismatch reflects an affective bias: the frames contain words that, according to an external database, would elicit trust, yet the participants perceive them as highly negative.

\paragraph{Concreteness analysis results}
Regarding the concreteness scores, among both high schoolers and psychology undergraduates, the frame of \textit{Grade} is significantly less concrete than random (H-S L-anx: $\bar{x}-\hat{\mu}_0 = -0.31$, $Z = -1.96$, Cohen's $|d| = 0.34$ Cliff's $|\delta| = 0.18$; Psy L-anx: $\bar{x}-\hat{\mu}_0 -0.24$, $Z = -2.40$, Cohen's $|d| = 0.25$, Cliff's $|\delta| = 0.13$, Psy H-anx  $\bar{x}-\hat{\mu}_0 = -0.21$, $Z = -2.22$, Cohen's $|d| = 0.22$, Cliff's $|\delta| = 0.11$), implying that grades are framed more as abstract evaluative labels (e.g., expectations, averages) than as directly interpretable outcomes. Interestingly, these significant concreteness results contrast with those found for the \textit{exam} frame, where the experts' semantic frame appeared more concrete than a random baseline. These findings, together with the difference in the valence aura of the respective samples in both concepts (i.e., \textit{exam} being framed neutrally by the experts and \textit{grade} being framed negatively by high school and psychology students), support the idea that abstractness may contribute to the sense of anxiety and loss of control that surrounds evaluation.

\begin{table}[!hbpt]
\scriptsize
\centering
\begin{tabular}{lccccp{8cm}}
\toprule
Sample & $N_v$ & $N_e$ & $C_i$ & $l_G$ & Hubs (top 5\% of degree distribution) \\
\midrule
Experts & 7 & 7 & 0.07 & 1.67 & Grade (6) \\
H-S South & 35 & 78 & 0.08 & 1.87 & Grade (34), \textit{Mathematics} (10) \\
H-Anx H-S & 59 & 110 & 0.03 & 1.94 & \textit{Grade} (58), \textit{Anxiety} (14), \textit{Exam} (13) \\
L-Anx H-S & 47 & 68 & 0.02 & 1.94 & \textit{Grade} (46), \textit{Anxiety} (9), \textit{Future} (8) \\
Physics & 22 & 29 & 0.04 & 1.87 & \textit{Grade} (21), \textit{Exam} (5), \textit{Anxiety} (5) \\
H-Anx Psy & 114 & 506 & 0.06 & 1.92 & \textit{Grade} (113), \textit{Anxiety} (41), \textit{Exam} (36), \textit{Task} (33), \textit{School} (29), Degree (28) \\
L-Anx Psy & 125 & 464 & 0.04 & 1.94 & \textit{Grade} (124), \textit{Exam} (38), \textit{Challenge} (33), \textit{University} (28), \textit{Anxiety} (28), \textit{Exercise} (27), \textit{School} (27) \\
\bottomrule
\end{tabular}
\caption{Network features for the semantic frame of \textit{\textbf{Grade}}.}
\label{tab:network_features_grade}
\end{table}

\subsection{Anxiety: behavioural forma mentis networks} \label{anxiety}
After exploring the concepts of \textit{exam} and \textit{grade}, which are the core of the evaluation process in the academic context, we now move on to the analysis of \textit{anxiety} and \textit{wellbeing} frames. In this case, we mainly focused on psychology undergraduates' samples, since they present the richest semantic frames for these cue words. We aim to grasp how \textit{anxiety} and \textit{wellbeing} are conceptualised in terms of semantic associations and emotional valence. Furthermore, we focus on how these concepts might be associated with the evaluation process and the academic context, and the potential differences between humans and GPT samples. To do so, we examine the similarity between the \textit{anxiety} and \textit{exam} semantic frames using the Jaccard Index, which allowed us to quantify the strength of associations among these concepts. Finally, as we did with the previous semantic frames, we present the differences in the concreteness ratings of nodes associated with \textit{anxiety} and \textit{wellbeing}, to assess whether these concepts are represented more concretely or abstractly across groups.

\begin{figure}[!hbpt]
    \centering
    \includegraphics[width=0.85\linewidth]{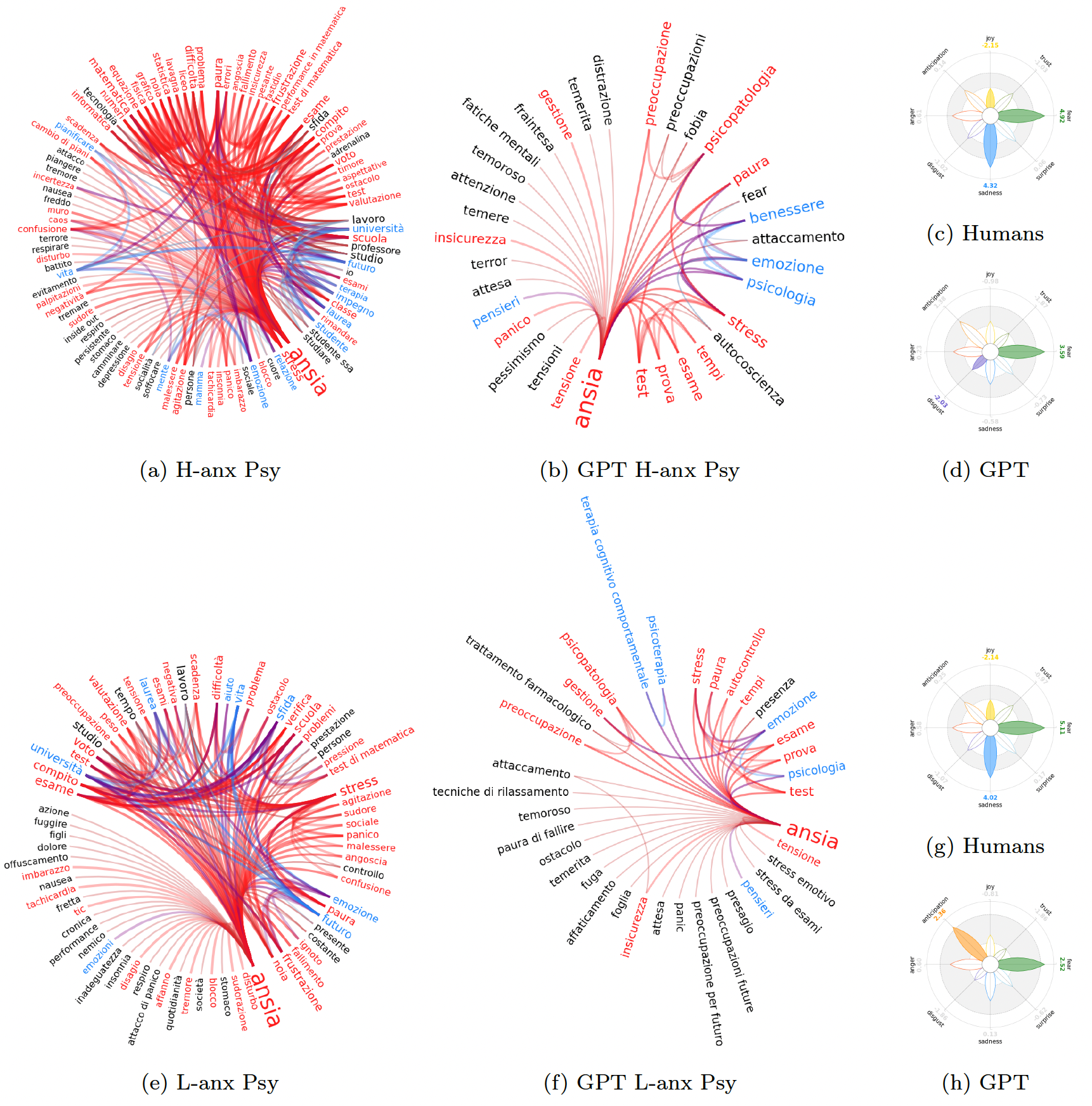}
    \caption{Semantic frames of the node \textbf{\textit{Anxiety}}.}
    \label{fig:anxiety_semantic_frames}
\end{figure}

\paragraph{Human vs. GPT semantic framing}
Figure \ref{fig:anxiety_semantic_frames} shows a clear divergence in how humans and GPT frame the concept of \textit{anxiety}. Human participants predominantly associate anxiety with concrete academic triggers, such as "school", "exam", and "grade". In contrast, GPT models link anxiety to more abstract and diagnostic concepts, including "panic", "emotion", and "stress". Several of these abstract terms also appear among the most central nodes in GPT and psychology-related samples (Table \ref{tab:network_features_anxiety}), reinforcing this pattern. These structural differences are mirrored by the concreteness analysis. GPT semantic frames are consistently less concrete than expected by chance (GPT H-S L-anx: $\bar{x}-\hat{\mu}_0 = -0.55$, $Z = -1.85$, Cohen’s $|d| = 0.56$, Cliff’s $|\delta| = 0.28$; GPT Psy H-anx: $\bar{x}-\hat{\mu}_0 = -0.49$, $Z = -2.26$, Cohen’s $|d| = 0.23$; GPT Psy L-anx: $\bar{x}-\hat{\mu}_0 = -0.32$, $Z = -1.65$, Cohen’s $|d| = 0.15$; Psy L-anx: $\bar{x}-\hat{\mu}_0 = -0.13$, $Z = -1.65$, Cohen’s $|d| = 0.04$), whereas human frames do not show this systematic abstraction. Taken together, these results suggest that GPT encodes anxiety primarily through clinical or theoretical terminology, while human students ground it in everyday academic situations.

\paragraph{Concrete anchors in human samples}
Southern high school students display a significantly positive concreteness score for \textit{anxiety} ($\bar{x} - \hat{\mu}_0 = 0.35$, $Z = 1.72$, Cohen’s $|d| = 0.37$, Cliff’s $|\delta| = 0.25$). This indicates that even an internal emotional state is represented through specific situational anchors, including exams, grades, professors, and time constraints. Notably, many of these concrete nodes are associated with negative valence, underscoring the role of academic actors, places, and tasks as sources of anxiety. By contrast, GPT low-anxiety high schoolers (\textit{GPT H-S L-anx}) and GPT psychology groups rely on markedly abstract and diagnostic representations, such as \textit{psychopathology} and \textit{stress}, with little reference to lived or situational experiences. This further highlights the divergence between experiential and classificatory modes of representation.

\begin{figure}[!hbpt]
    \centering
    \includegraphics[width=0.6\textwidth]{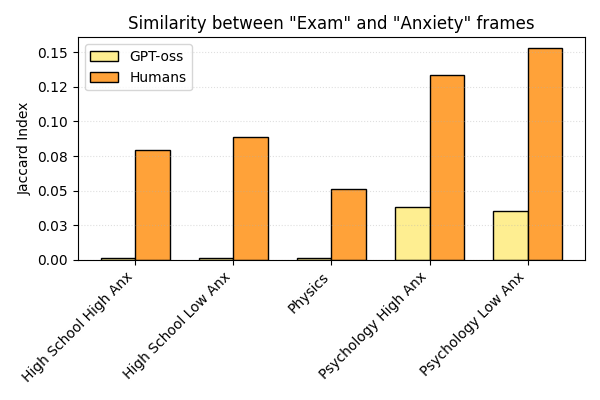}
    \caption{Distribution of Jaccard similarity values between semantic frames of \textit{Exam} VS \textit{Anxiety} across samples. Values equal to zero were plotted at $J = 0.001$ for visualisation purposes, so that zero-overlap cases appear as a visible bar.}
    \label{fig:exam_anxiety_comparison}
\end{figure}

\paragraph{Overlap with academic concepts}
Semantic overlap between \textit{anxiety} and \textit{exam} was quantified using the Jaccard Index. As shown in Figure \ref{fig:exam_anxiety_comparison}, human semantic networks exhibit substantially higher similarity than GPT-oss networks, with the degree of overlap modulated by anxiety level. In high-anxiety human groups, Jaccard similarity between \textit{exam} and \textit{anxiety} frames exceeds 0.13, approximately three to four times higher than values observed in GPT-oss networks, which rarely surpass 0.04. These findings indicate that, in the human semantic organisation, anxiety is tightly embedded within academic concept networks. In contrast, GPT representations tend to keep \textit{anxiety} and \textit{exam} largely segregated, treating them as conceptually adjacent but structurally distinct.

\paragraph{Emotional patterns}
Turning to emotional profiles, no major differences emerge across groups. Fear is significantly represented in all samples (H-anx Psy: $z = 4.92$; GPT H-anx Psy: $z = 3.59$; L-anx Psy: $z = 5.11$; GPT L-anx Psy: $z = 2.52$). However, a significant presence of sadness is observed only in human samples (H-anx Psy: $z = 4.32$; L-anx Psy: $z = 4.02$), suggesting that human representations of anxiety capture a broader affective spectrum than those generated by GPT.

\begin{table}[!hbpt]
\scriptsize
\centering
\begin{tabular}{lccccp{8cm}}
\toprule
Sample & $N_v$ & $N_e$ & $C_i$ & $l_G$ & Hubs (top 5\% of degree distribution) \\
\midrule
H-S South & 24 & 55 & 0.13 & 1.80 & Anxiety (23), Grade (9) \\
H-Anx H-S & 75 & 218 & 0.05 & 1.92 & \textit{Anxiety} (74), \textit{Stress} (23), \textit{Fear} (21), \textit{School} (19) \\
L-Anx H-S & 51 & 110 & 0.05 & 1.91 & \textit{Anxiety} (50), \textit{Stress} (18), \textit{Fear} (11), \textit{Embarrassment} (11) \\
GPT H-Anx H-S & 14 & 17 & 0.05 & 1.81 & \textit{Anxiety} (13) \\
GPT L-Anx H-S & 12 & 13 & 0.04 & 1.80 & \textit{Anxiety} (11) \\
Physics & 31 & 64 & 0.08 & 1.86 & \textit{Anxiety} (30), \textit{Stress} (9) \\
GPT Physics & 7 & 8 & 0.13 & 1.62 & \textit{Anxiety} (6) \\
H-Anx Psy & 253 & 1376 & 0.04 & 1.96 & \textit{Anxiety} (252), \textit{Stress} (69), Work (53), \textit{Boredom} (52), \textit{Exam} (51), \textit{School} (50), \textit{Frustration} (49), \textit{Challenge} (46), \textit{Mathematics} (45), \textit{Study} (41), \textit{Grade} (41), \textit{Emotion} (40), \textit{Task} (40), \textit{University} (40) \\
L-Anx Psy & 223 & 944 & 0.03 & 1.96 & \textit{Anxiety} (222), \textit{Stress} (63), \textit{Exam} (47), \textit{Challenge} (46), \textit{University} (44), Work (38), \textit{School} (36), \textit{Frustration} (34), \textit{Future} (32), \textit{Psychology} (32), \textit{Task} (32), \textit{Study} (32), \textit{Boredom} (32) \\
GPT H-Anx Psy & 33 & 57 & 0.05 & 1.89 & \textit{Anxiety} (32), \textit{Stress} (8) \\
GPT L-Anx Psy & 36 & 55 & 0.03 & 1.91 & \textit{Anxiety} (35), Tests (6), \textit{Exam} (6), \textit{Stress} (6) \\
\bottomrule
\end{tabular}
\caption{Network features for the semantic frame of \textit{\textbf{Anxiety}}.}
\label{tab:network_features_anxiety}
\end{table}

\subsection{Wellbeing: behavioural forma mentis networks}
\begin{figure}[!hbpt]
    \centering
    \includegraphics[width=0.75\linewidth]{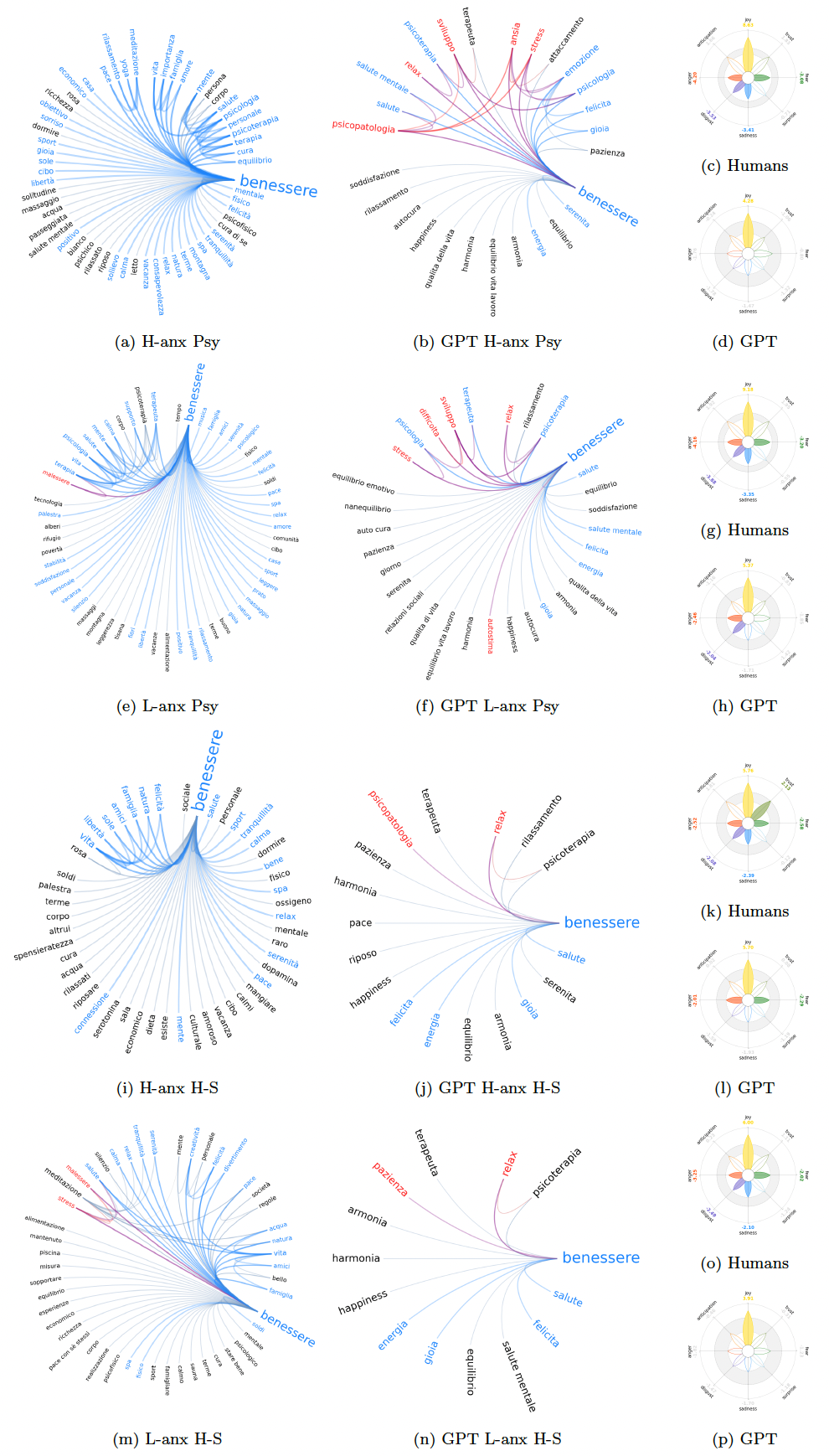}
    \caption{Semantic frames of the node \textbf{\textit{Wellbeing}}.}
    \label{fig:wellbeing_semantic_frames}
\end{figure}

As with the \textit{anxiety} frame, we analyse the semantic organisation of \textit{wellbeing} by comparing human and GPT samples, as well as differences across human groups (e.g., high school and psychology students). Figure \ref{fig:wellbeing_semantic_frames} illustrates the \textit{wellbeing} semantic frames for each sample, while Table \ref{tab:network_features_wellbeing} reports their associated network features.

\paragraph{Wellbeing and therapy}
Figure \ref{fig:wellbeing_semantic_frames} highlights marked differences in how \textit{wellbeing} is framed by humans and GPT-simulated students. GPT models tend to associate \textit{wellbeing} with negatively valenced psychological constructs, such as \textit{psychopathology}, \textit{anxiety}, and \textit{stress}, alongside explicit references to clinical interventions (e.g., \textit{therapy}, \textit{psychotherapy}). In contrast, human participants, particularly high school and psychology students, primarily link \textit{wellbeing} to concrete, everyday practices, including \textit{diet}, \textit{sleep}, and \textit{physical activity}. Notably, references to therapy are absent in high school and physics student samples, suggesting differences in psychoeducational exposure and conceptual focus across groups.

\paragraph{Concreteness of wellbeing representations}
The concreteness analysis supports these qualitative differences. Southern high school students exhibit a significantly positive concreteness score for \textit{wellbeing} ($\bar{x} - \hat{\mu}_0 = 0.43$, $Z = 2.10$, Cohen’s $|d| = 0.44$, Cliff’s $|\delta| = 0.28$), indicating that wellbeing is represented through specific, everyday experiences rather than abstract psychological constructs. By contrast, GPT-simulated psychology students with high anxiety show a negative deviation from the random baseline ($\bar{x} - \hat{\mu}_0 = -0.34$, Cohen’s $|d| = -1.68$, Cliff’s $|\delta| = 0.15$), reflecting a more abstract framing. Overall, these patterns mirror the semantic-frame results, with GPT representations emphasising diagnostic and clinical concepts, while human representations prioritise lifestyle factors and daily behaviours.

\begin{table}[!hbpt]
\scriptsize
\centering
\begin{tabular}{lccccp{8cm}}
\toprule
Sample & $N_v$ & $N_e$ & $C_i$ & $l_G$ & Hubs (top 5\% of degree distribution) \\
\midrule
Experts & 11 & 12 & 0.04 & 1.78 & Wellbeing (10) \\
H-S South & 27 & 37 & 0.03 & 1.89 & Wellbeing (26), Health (6) \\
H-Anx H-S & 49 & 55 & 0.01 & 1.95 & \textit{Wellbeing} (48), Life (8), Sun (2), Pink (2), Nature (2), Friends (2), Freedom (2), Happiness (2), Family (2) \\
L-Anx H-S & 49 & 74 & 0.02 & 1.94 & \textit{Wellbeing} (48), \textit{Meditation} (12), Life (8) \\
GPT H-Anx H-S & 18 & 19 & 0.01 & 1.88 & \textit{Wellbeing} (17) \\
GPT L-Anx H-S & 14 & 14 & 0.01 & 1.85 & \textit{Wellbeing} (13) \\
Physics & 26 & 25 & 0.00 & 1.92 & \textit{Wellbeing} (25), Flowers (1), Sun (1), Awareness (1), Tranquility (1), Love (1), Spa (1), Green (1), Serenity (1), Time (1), Health (1), Presence (1), Support (1), Holiday (1), Water (1), Affection (1), Psychophysical (1), Relax (1), Family (1), Money (1), Balance (1), Calm down (1), Peace (1), Society (1), Sea (1), Security (1) \\
GPT Physics & 11 & 10 & 0.00 & 1.82 & \textit{Wellbeing} (10) \\
H-Anx Psy & 141 & 337 & 0.02 & 1.97 & \textit{Wellbeing} (140), \textit{Emotion} (28), \textit{Psychology} (23), \textit{Mind} (20), \textit{Discovery} (20), \textit{Future} (19), \textit{Passion} (18), \textit{Fun} (18) \\
L-Anx Psy & 148 & 429 & 0.03 & 1.96 & \textit{Wellbeing} (147), Work (29), \textit{Future} (27), \textit{Psychology} (26), \textit{Emotion} (26), \textit{Mind} (22), \textit{Stress} (22), Life (22) \\
GPT H-Anx Psy & 27 & 45 & 0.06 & 1.87 & \textit{Wellbeing} (26), \textit{Emotion} (7) \\
GPT L-Anx Psy & 31 & 37 & 0.02 & 1.92 & \textit{Wellbeing} (30), \textit{Development} (5) \\
\bottomrule
\end{tabular}
\caption{Network features for the semantic frame of \textit{\textbf{Wellbeing}}.}
\label{tab:network_features_wellbeing}
\end{table}

\FloatBarrier

\section{Discussion}
This study adopted a cognitive network science approach to examine how different academic populations mentally organise test anxiety-related concepts, using behavioural forma mentis networks \cite{stella2019forma}. By integrating free associations with group-level valence ratings, semantic frame structure, and psycholinguistic grounding through concreteness norms \cite{brysbaert2014concreteness}, we investigated whether distinct profiles of math anxiety correspond to systematic differences in the affective and conceptual organisation of the academic evaluation process. Human networks were further compared with GPT-oss simulated students, prompted to approximate comparable anxiety and educational profiles.

In this Section, we discuss the results in relation to the three research questions outlined in the Introduction. We examine how academic populations frame evaluative concepts such as \textit{exams} and \textit{grades} across educational stages and anxiety profiles, focusing on differences in semantic organisation and affective tone. Then, we analyse how evaluation-related concepts vary across groups in terms of network structure, valence aura, and concreteness, with particular attention to patterns associated with high anxiety. Finally, we compare human semantic frames with GPT-oss–based networks to assess the extent to which artificial systems reproduce human cognitive and emotional patterns of test anxiety.

\subsection{Evaluative concepts as affectively charged and experientially distant}
% RQ1 – How do academic populations frame evaluative concepts (e.g., exams, grades) across educational stages and anxiety profiles?
From our analysis of the semantic frame of \textit{exam} and \textit{grade}, a strong negative framing persists across all academic populations examined here. Instead, experts seem to have a more distant and neutral perception of these concepts. Semantic frames consistently cluster \textit{exam} and \textit{grade} around notions of \textit{failure}, \textit{pressure}, \textit{evaluation}, and \textit{anxiety}, suggesting that academic evaluation is widely framed and experienced as a crucial judgement rather than as a formative learning opportunity \cite{putwain2008deconstructing}. This pattern holds across educational stages and anxiety profiles, highlighting the pervasiveness of evaluative stress within academic cultures and supporting previous research \cite{foley2017mathanxiety, luttenberger2018spotlight}.

Emotional analyses reinforce this interpretation. Fear and anticipation were repeatedly associated with \textit{exam}, while \textit{grade} frames elicit a complex affective mix in which negative and neutral associations coexist with a statistically significant presence of trust. This contrast represents an emotional bias \cite{DeDuro2025}, suggesting that evaluative concepts activate institutional or normative expectations of fairness and legitimacy, even when they are subjectively experienced as threatening. Such affective dissonance may contribute to the persistence of test anxiety \cite{carver1983effects}, as students simultaneously recognise evaluation as necessary or legitimate while experiencing it as emotionally burdensome.

Notably, this negative framing is not limited to student populations. GPT-simulated samples reproduce and, in some cases, amplify this pervasive negativity, indicating that such representations reflect widespread human experiences about the academic evaluation system rather than idiosyncratic experiences. However, important differences between GPT and humans emerge when considering the concrete, experience-related component of these framings.

\subsection{Concreteness and perceived controllability}
% RQ2 – How do evaluation-related concepts differ in network structure, valence aura, and concreteness across groups?
A central structural distinction across groups concerns the concreteness with which evaluative concepts are represented. Our BFMNs analysis suggests that \textit{grade} tends to be framed as an abstract evaluative construct, associated with labels such as expectations, averages, and judgment, rather than as a directly interpretable outcome tied to specific actions or performances. This abstraction is particularly pronounced in high-anxiety groups, especially among high school and psychology student samples. 

In contrast, the experts' semantic frame of \textit{exam} stands out as both emotionally neutral and significantly more concrete. Within this population, exams are represented through tasks, procedures, and outcomes rather than through threat-related evaluative meanings. This divergence suggests that experiential familiarity and professionalisation may restore concreteness to evaluative concepts, thereby reducing their emotional charge. Importantly, the dissociation between \textit{exam} and \textit{grade} is theoretically meaningful: while exams can become proceduralised and grounded through repeated exposure, grades remain abstract symbols of evaluation, even for advanced students. This persistent abstraction may help explain why grades continue to elicit anxiety despite increased academic expertise.

These patterns can be interpreted through cognitive theories of anxiety, which emphasise biased threat processing and reduced cognitive control. In the cognitive model proposed by Beck at al, 1997 \cite{beckInformationProcessingModel1997}, anxiety is fundamentally characterised by the selective processing of information perceived as threatening. The intermediate stage of processing involves the activation of a primal threat mode, characterised by narrowing of cognitive processing, rigid thinking, and intolerance for uncertainty or ambiguity \cite{beckInformationProcessingModel1997, clarkCognitiveTheoryTherapy2010}. Our finding that evaluation-related concepts such as \textit{Grade} are significantly less concrete than a random baseline in high-anxiety groups is consistent with this cognitive constriction. Specifically, the shift away from concrete, actionable representations toward abstract and global framings may amplify psychological distance and reinforce perceived uncontrollability. This interpretation is also coherent with cognitive models of generalised anxiety disorder \cite{wellsCognitiveModelGeneralized1999}, where negative metabeliefs sustain worry by focusing on the perceived uncontrollability and dangerous consequences of worrying. When evaluative concepts are framed abstractly, their lack of clear behavioural anchors may strengthen the perception that the individual has limited control over outcomes, thereby fuelling \textit{Type 2} worry, a defining cognitive feature of anxiety \cite{beckInformationProcessingModel1997, spielberger1970manual}.

Taken together, these findings indicate that differences in evaluative framing across groups are not limited to emotional valence but extend to the structural and experiential organisation of semantic representations. More broadly, abstraction may represent a key cognitive mechanism through which academic evaluation becomes psychologically threatening, particularly in individuals with higher anxiety.

\subsection{Human and GPT semantic framing of test anxiety and wellbeing}
% RQ3 – How do these human mindsets compare with corresponding GPT-oss–based networks, and what does this reveal about modelling test anxiety in artificial systems?

The analysis of the \textit{anxiety} frame clarifies how test anxiety and evaluative stress are integrated into academic meaning systems. Human participants consistently anchor anxiety to concrete academic triggers, including \textit{exams}, \textit{grades}, \textit{school}, and time constraints. These associations are particularly common in high-anxiety groups and are reflected in higher semantic overlap between the \textit{anxiety} and \textit{exam} frames. Such overlap indicates that, in human semantic organisation, anxiety is not a detached emotional state but a structurally embedded component of academic evaluation, closely tied to the lived experience of being assessed \cite{putwain2008deconstructing}.

This pattern should also be interpreted in light of the task design. Since the free-association paradigm relied on cue words drawn primarily from academic life, human participants were naturally encouraged to retrieve context-specific experiences. As a result, the strong anchoring of \textit{anxiety} to academic triggers partly reflects the fact that participants were operating within a familiar semantic domain shaped by repeated exposure to evaluative situations. 

However, this methodological constraint applies equally to GPT-oss simulations, which were exposed to the same academic cue set and prompted with comparable student profiles. Despite this shared framing, GPT networks did not reproduce the same contextual embedding of anxiety into evaluative academic concepts. Instead, GPT-oss responses remained centred on abstract, diagnostic, and clinical terminology (e.g., \textit{stress}, \textit{panic}, \textit{psychopathology}), and the overlap between \textit{anxiety} and \textit{exam} remained close to null. This suggests that the divergence cannot be explained solely by cue-word priming, but reflects a bigger difference in how human and artificial systems organise emotional meaning: humans retrieve test anxiety through lived academic contexts, whereas GPT encodes anxiety primarily as decontextualised semantic knowledge \cite{wellsCognitiveModelGeneralized1999}.

At the same time, GPT-oss simulations successfully reproduced some large-scale semantic regularities observed in human data, including the pervasive negative framing of evaluation-related concepts (i.e., \textit{grade} and \textit{exam}). This convergence indicates that LLMs can approximate aggregate cultural narratives about academic assessment and evaluative pressure when prompted with plausible educational profiles, consistent with previous work showing that LLMs internalise population-level semantic patterns \cite{abramski2023cognitive, ciringione2025math}. However, the absence of strong coupling between \textit{anxiety} and core evaluative concepts suggests that GPT captures what students tend to say about anxiety at a discursive level, but not how test anxiety is cognitively organised as an experience-dependent mindset.

The semantic framing of \textit{wellbeing} reveals a complementary contrast. Human participants, particularly high school and psychology students, define \textit{wellbeing} primarily through concrete lifestyle practices such as \textit{sleep}, \textit{diet}, and \textit{physical activity}. GPT-simulated students, on the other hand, frame \textit{wellbeing} through negatively valenced psychological and therapeutic constructs. This divergence further underscores the role of lived experience in shaping mental health representations: while GPT models reproduce diagnostic and institutional discourses surrounding wellbeing, they fail to capture the everyday practices through which individuals experience and regulate it. Together with the anxiety results, this pattern highlights a systematic gap between semantic generalisation and experiential grounding in artificial systems \cite{ciringione2025math}.

Overall, these findings reinforce the view of test anxiety as a situated, experience-dependent phenomenon \cite{putwain2008deconstructing}, and they demonstrate both the strengths and limitations of GPT-based simulations in modelling academic emotional mindsets.

\subsection{Future implications and limitations}

These findings suggest that, even in STEM samples, test anxiety is not merely an attitudinal stance toward evaluation, but a structurally embedded and experientially grounded mindset shaped by repeated interactions with academic assessment. From an applied perspective, the broad negativity framing of concepts related to the academic evaluation systems suggests the implementation of different evaluative methods and psychological support to help students cope with test anxiety. Furthermore, the abstraction of evaluative concepts in high-anxiety groups suggests that interventions may benefit from increasing perceived controllability by grounding evaluation in clearer, more concrete and actionable representations (e.g., transparent criteria and formative feedback). Behavioural forma mentis networks could also provide a quantitative framework for monitoring whether interventions reshape semantic frames toward less threatening configurations. 

Concerning GPT-oss, the abstract framing of test anxiety-related concepts suggests that LLM simulations may capture what students tend to say about evaluation-related stress, but not how such stress is cognitively embedded in lived experience \cite{ciringione2025math}. As a consequence, GPT-based outputs should not be treated as direct psychological proxies without careful validation.

Despite the important practical implications of these findings, our study still presents some limitations. Firstly, our samples differ in size and in the materials they received (i.e., the set of cue words and the MAS-IT questionnaire). Furthermore, all the human participants were recruited through convenience sampling, which does not ensure a representative sample of the population we aimed to study. All of these aspects limit generalisability and comparability. Secondly, since the study is cross-sectional, we cannot draw causal conclusions, and longitudinal designs are needed to test whether semantic structure predicts test anxiety or whether anxiety reshapes semantic representations over time. Future work should address these limitations by adopting longitudinal and intervention-based designs to test whether changes in semantic framing precede or follow changes in test anxiety. In addition, linking network measures to behavioural outcomes (e.g., exam performance, course choices, or dropout risk) would clarify their predictive value. Finally, extending recruitment to more diverse and balanced academic populations would strengthen the generalisability of these findings.

\section{Conclusion}
Overall, our results show that academic evaluation is consistently framed as emotionally threatening across student populations, with \textit{exam} and \textit{grade} embedded in negative semantic neighbourhoods. High-anxiety groups additionally represent evaluative concepts in more abstract terms, suggesting reduced perceived controllability and increased psychological distance from assessment outcomes. Human semantic networks also reveal that test anxiety is strongly grounded in concrete academic experiences, whereas GPT-oss simulations reproduce cultural negativity but fail to embed anxiety within academic contexts. Taken together, these findings support the view of test anxiety as a structurally embedded and experience-dependent mindset, and highlight both the potential and the limits of LLMs as tools for modelling academic emotional cognition.us

\section*{Data Availability Statement}
All the collected data used for this study are available on an OSF repository at the following link: \url{https://doi.org/10.17605/OSF.IO/FTGSV}.

\section*{Acknowledgements}
We acknowledge support from Fondazione Caritro (Grant RASSERENO).

\section*{Declarations}
We have no known conflict of interest to disclose.

\section*{Author contributions}
% Conceptualisation, Methodology, Validation, Formal analysis, Investigation, Data curation: E.F., F.G., M.S.; Visualisation: E.F., F.G; Writing – original draft: E.F., F.G., M.S.; Writing – review \& editing: E.F., F.G., M.S., A.G., G.L.; Supervision, Project administration: M.S., G.L.

\bibliographystyle{iopart-num}  
\bibliography{bibliography.bib}

\end{document}